\documentclass[12pt]{amsart}
\usepackage{amsbsy,amssymb,amscd,amsfonts,latexsym,amstext,delarray,
amsmath,graphicx} 
\input xypic

\usepackage{color}

\numberwithin{equation}{section}

\def\bK{{\mathbb K}}
\def\bL{{\mathbb L}}

\def\A{{\mathbb A}}
\def\C{{\mathbb C}}
\def\F{{\mathbb F}}
\renewcommand{\H}{{\mathbb H}}

\renewcommand{\P}{{\mathbb P}}
\def\Q{{\mathbb Q}}
\def\R{{\mathbb R}}
\def\Z{{\mathbb Z}}
\def\K{{\mathbb K}}

\def\m{{\mathfrak m}}

\def\cC{{\mathcal C}}
\def\cD{{\mathcal D}}
\def\cE{{\mathcal E}}
\def\cF{{\mathcal F}}
\def\cG{{\mathcal G}}
\def\cH{{\mathcal H}}

\def\cL{{\mathcal L}}

\def\cO{{\mathcal O}}
\def\cP{{\mathcal P}}

\def\cR{{\mathcal R}}
\def\cS{{\mathcal S}}
\def\cT{{\mathcal T}}

\def\Aut{{\rm Aut}}

\def\GL{{\rm GL}}

\def\PGL{{\rm PGL}}
\def\PSL{{\rm PSL}}

\def\SL{{\rm SL}}

\def\Tr{{\rm Tr}}

\title[Holographic Codes on BT Buildings and Drinfeld Spaces]{Holographic Codes on 
Bruhat--Tits buildings and Drinfeld Symmetric Spaces}
\author{Matilde Marcolli}
\address{Department of Mathematics, University of Toronto, Canada  \newline  \indent 
Perimeter Institute for Theoretical Physics, Waterloo, Canada \newline  \indent 
Division of Physics, Mathematics, and Astronomy, Caltech, USA}
\email{matilde@caltech.edu}

\begin{document}
\maketitle

\begin{abstract}
This paper is based on the author's talk at the Arbeitstagung 2017. It 
discusses some general approaches to the construction of classical and
quantum holographic codes on Bruhat--Tits trees and
buildings and on Drinfeld symmetric spaces, in the context of the 
$p$-adic AdS/CFT correspondence.
\end{abstract}

\begin{verse}
{\em Dedicated to Yuri Manin on the occasion of his 80th birthday}
\end{verse}

\section{Introduction} 

This paper is based on the talk given by the author at the Arbeitstagung 2017
``Physical Mathematics" in honor of Yuri Manin's 80th birthday. 
It is an introduction to an ongoing joint project with Matthew Heydeman,
Sarthak Parikh and Ingmar Saberi, on the construction of holographic classical and
quantum codes on Bruhat--Tits trees and higher rank Bruhat--Tits buildings and on
Drinfeld symmetric spaces, and associated entanglement entropy formulae. A discussion
of the entanglement entropy and the relation to other holographic codes constructions,
such as \cite{HaPPY}, will be presented in a forthcoming joint paper in preparation. The present paper
should be regarded as covering some background material on the question of constructing
holographic codes on $p$-adic symmetric spaces, based on algebro-geometric properties. 

\smallskip

In \cite{Manin1}, \cite{Manin2}, Manin gave a compelling view of the idea of ``Arithmetical
Physics", according to which physics in the usual Archimedean setting or real and complex numbers
would cast non-Archimedean shadows that live over the finite primes and arithmetic properties 
associated to these non-Archimedean models can be used to better understand the physics that we
experience at the Archimedean ``prime at infinity". According to this general philosophy
${\rm Spec}(\Z)$ is the ``arithmetic coordinate" of physics and geometry. A famous example
where this principle manifests itself is given by the description of the Polyakov measure 
for the bosonic string in terms of the Faltings height function at algebraic points of the 
moduli space of curves, which leads naturally to the question of whether the Polyakov
measure is in fact an adelic object and whether there is an overall arithmetic expression
for the string partition function, \cite{Manin3}, \cite{Manin4}. More generally, one can
ask to what extent are the fundamental laws of physics adelic. 
Does physics in the Archimedean setting (partition functions, action functionals, real 
and complex variables) have $p$-adic manifestations? Can these be used to 
provide convenient ``discretized models? of physics, powerful enough to determine their 
Archimedean counterpart?

\smallskip

Various forms of $p$-adic and adelic phenomena in physics and their
relation to the usual Archimedean formulation were developed over the years.
We refer the readers to \cite{CMZ}, \cite{DKKV}, \cite{VVZ}, \cite{Zab} 
for some references relevant to the point of view discussed in this paper.

\smallskip

Here we focus in particular on the holographic AdS/CFT correspondence
and on the recent viewpoint relating information (entanglement entropy)
of quantum states on the boundary to geometry (classical gravity) on the
bulk, \cite{NRT} and the tensor networks and holographic codes approach
of \cite{HaPPY}.  The existence of a $p$-adic version of the holographic
AdS/CFT correspondence was already proposed in \cite{ManMar}, based
on earlier results of Manin \cite{Manin5}, \cite{Manin6} expressing the
Green function on a compact Riemann surface with Schottky uniformization
to configurations of geodesics in the bulk hyperbolic handlebody (which
are higher genus generalizations of Euclidean BTZ black holes \cite{Krasnov})
and results of Drinfeld and Manin \cite{ManDri} on periods of Mumford
curves uniformized by $p$-adic Schottly groups.

\smallskip

In \cite{HMSS} we developed a non-Archimedean version of AdS/CFT holography, based on the 
approach originally proposed in \cite{ManMar}, which would be compatible with the more recent
viewpoint on the holographic correspondence based on the ideas of tensor networks and holographic
codes and the correspondence between entanglement entropy and bulk geometry.
Versions of $p$-adic AdS/CFT correspondence were also developed in \cite{GKPSW}, and in subsequent work
\cite{BHLL}, \cite{GuPa}, \cite{GHJPSST}, \cite{GHJMPSST},  \cite{GJPT} and others. The theme of non-archimedean versions of holography has clearly become a very active area of current research. 

\smallskip

In this paper, we return to the point of view of tensor networks and holographic codes discussed in
\cite{HMSS} and we present some new constructions which are based on the geometry of Bruhat--Tits
trees and buildings and of Drinfeld symmetric spaces. 

\smallskip

The main difference between the approach we propose here and other
constructions of holographic codes such as \cite{HaPPY}, or for instance
\cite{BHLL}, \cite{BreRi}, \cite{HNQTWY}, lies in the fact that we rely on
well known techniques for the construction of classical codes associated
to algebro-geometric objects \cite{TsfaVla} and on algorithms relating classical to quantum
codes \cite{CRSS}. The construction of algebro-geometric codes played a crucial
role in the study of asymptotic problems in coding theory, as shown by Manin
in \cite{Manin7}.

\smallskip

We first present here a construction of holographic codes that is based on the geometry of
the Bruhat-Tits trees and algebro-geometric Reed--Solomon codes associated to 
projective lines over a finite field, together with an application of the CRSS algorithm
that associates quantum codes to classical $q$-ary codes. 

\smallskip

We then revisit the approach to holographic codes via tessellations of the hyperbolic plane,
as in \cite{HaPPY}. Instead of relating such constructions to the Bruhat--Tits trees via a
non-canonical planar embedding of the tree, as in \cite{HMSS}, we use here a purely
$p$-adic viewpoint, working with the Drinfeld $p$-adic upper half plane as a replacement
of the real hyperbolic plane, and its (canonical) map to the  Bruhat--Tits tree. Instead of
tessellations of the real hyperbolic plane we use actions of $p$-adic Fuchsian groups on
the Drinfeld plane and associated surface codes. We show that this approach is restricted 
by the strong constraints that exist on $p$-adic Fuchsian groups. For example, we show that
a $p$-adic analog of the holographic pentagon code of \cite{HaPPY} constructed with this
method can only exist when $p=2$.

\smallskip

We then propose an extension of this approach via holographic codes to higher rank buildings,
based on algebro-geometric codes associated to higher dimensional algebraic varieties, as
constructed in \cite{TsfaVla}.

\medskip
\section{Algebro-Geometric Codes on the Bruhat--Tits tree}\label{BTcodes}

In this section we describe a construction of holographic codes on the Bruhat--Tits trees
that are obtained via Reed--Solomon algebro-geometric codes on projective lines over finite fields.

\subsection{Reed--Solomon codes and classical codes on the Bruhat--Tits tree}

The set of algebraic points $X(\F_q)$ of a curve $X$ over a finite field $\F_q$ can be
used to construct algebro-geometric error-correcting codes, see \cite{TsfaVlaNo}. 
Algebro-geometric codes associated to a curve $X$ over a finite field $\F_q$
consists of a choice of a set $A$ of algebraic points $A\subset X(\F_q)$ and a divisor
$D$ on $X$ with support disjoint from $A$. The linear code $C=C_X(A,D)$ is obtained
by considering rational functions $f\in \F_q(X)$ with poles at $D$ and evaluating
them at the points of $A$. A bound on the order of pole of $f$ at $D$ determines
the dimension of the linear code.

\smallskip

We are interested here in the simplest case of algebro-geometric codes, the 
Reed--Solomon codes constructed using the points of $\P^1(\F_q)$.
Given a set of points $A \subset \P^1(\F_q)$ with $\# A =n \leq q+1$ we consider
two types of Reed--Solomon codes, one constructed using the point $\infty \in \P^1(\F_q)$ 
as divisor, that is, using polynomials $f\in \F_q[x]$, and using a set $A$ of $n\leq q$ points
in $\A^1(\F_q)=\F_q$ for evaluation. The corresponding Reed--Solomon code
$C=\{ (f(x_1),\cdots, f(x_n))\,:\, f\in \F_q[x], \, \deg(f)< k \}$ gives an $[n,k,n-k+1]_q$
classical code, where $n\leq q$. The other type of Reed--Solomon codes are obtained 
using homogeneous polynomials and a set $A$ of $n\leq q+1$ points in $\P^1(\F_q)$.
The resulting code $\hat C=\{ (f(u_1,v_1),\ldots,f(u_n,v_n))\,:\, f\in \F_q[u,v], \, \text{homogeneous with }
\deg(f)<k\}$, with $x_i=(u_i:v_i)\in \P^1(\F_q)$. We also consider generalized Reed-Solomon codes
of these two types, where for a vector $w=(w_1,\ldots,w_n)\in \F_q^n$ one defines
$$ C_{w,k}=\{ (w_1 f(x_1),\cdots, w_n f(x_n))\,:\, f\in \F_q[x], \, \deg(f)< k \} $$
$$ \hat C_{w,k}=\{ (w_1 f(u_1,v_1),\ldots, w_n f(u_n,v_n))\,:\, f\in \F_q[u,v], \, \text{homogeneous}, \,
\deg(f)<k\}. $$

\smallskip

For $\K$ a finite extension of $\Q_p$ with residue field $\F_q$, with $q=p^r$, the Bruhat--Tits tree
$\cT_\K$ is a homogeneous tree with valence $q+1=\# \P^1(\F_q)$ and with ends
$\partial \cT_\K =\P^1(\K)$. The choice of a projective coordinate on $\P^1(\K)$
fixes three points $\{0,1,\infty\} \in \P^1(\K)$, hence it fixes a unique root vertex 
$\nu_0\in V(\cT_\K)$. The star of vertices surrounding $\nu_0$ can then be identified with
a copy of $\P^1(\F_q)$, which in algebro-geometric terms corresponds to the reduction
modulo the maximal ideal $\m$ in $\cO_\K$. 

\smallskip

The root vertex $\nu_0$ is therefore associated to the reduction curve $\P^1$. We can
construct a holographic {\em classical code} to the Bruhat--Tits tree by assigning
to the root vertex $\nu_0$ and its star of $q+1$ edges a Reed--Solomon code with
an assigned number $k$ of logical inputs ($q$-ary bits) located at $\nu_0$ and outputs 
at each of the $q+1$ legs. This can be done by a (generalized) Reed-Solomon code
$\hat C_{w,k}$ of maximal length $n=q+1$, seen as an encoding $\hat C_{w,k} : \F_q^k \to \F_q^{q+1}$,
which inputs a $k$-tuple of $q$-ary bits $a=(a_0,\ldots, a_{k-1})\in \F_q^k$, uses the
homogeneous polynomial $f_a(u,v)=\sum_{i=0}^k a_i u^iv^{k-1-i}$, and outputs a
$q$-ary bit $f(u_j,v_j)\in \F_q$ at each point $x_j=(u_j:v_j)\in \P^1(\F_q)$ identified with a
leg of the vertex $\nu_0$ in the Bruhat--Tits tree. 

\smallskip

The choice of the projective coordinate on $\P^1(\K)$, hence of the root vertex $\nu_0$ in
$\cT_\K$, determines a choice of a leg at each other vertex $\nu\neq \nu_0$, given by the
unique direction out of $\nu$ towards the root $\nu_0$. We can identify this choice with a
choice of the point $\{ \infty \}$ in each copy of $\P^1(\F_q)$ at each vertex $\nu\neq \nu_0$
of the tree. Proceeding from the center, if we assign a Reed--Solomon code at each
vertex of $\cT_\K$, and by homogeneity we expect all of them to have the same
number $k$ of inputs, we see that at each successive steps the leg of the star of
edges at $\nu$ has already one value assigned at the leg labelled by the point $\infty\in \P^1(\F_q)$,
which corresponds to the output coming from the matching leg in the star of the
previous vertex coming from the root $\nu_0$. Thus, in projective coordinates $(u:v)$ where
$(0:1)$ is the point at infinity, the Reed--Solomon code
$\hat C_{w,k}$ associated to the vertex $\nu$ takes $k-1$ new inputs $a=(a_1,\ldots,a_{k-1})\in \F_q^{k-1}$ 
and one additional input $a_0$ given by the value at $\infty$ assigned by the previous code, and 
deposits a new $q$-ary bit $f_a(u,v)=\sum_{i=0}^{k-1} a_i u^i v^{k-1-i}$ 
at each of the remaining legs at the vertex $v$ pointing away from the root, labelled by the points 
$x\in \A^1(\F_q)=\F_q$.

\smallskip

Note how the construction considered here has one root vertex play a special
role with an $\F_q^k$ logical input and a Reed--Solomon code of length $q+1$,
while all the other vertices have a further logical input of $\F_q^{k-1}$. 
This asymmetry is inevitable if we want
to use the algebro-geometric structure underlying the Bruhat--Tits tree to
construct a classical code, since the root vertex plays the special role of
the algebraic curve given by the reduction modulo $\m$, while the sets of
vertices in the tree at distance $m$ from the root correspond to reducing
modulo powers $\m^m$. Thus, the asymmetric role of the root vertex and
the other vertices is built into the relation between $\P^1(\K)$ and its
reduction curves.

\smallskip

The construction described here determines a classical code associated to the Bruhat--Tits tree with logical inputs 
at the vertices and outputs at the forward pointing legs. In the limit where one considers
the whole tree, the outputs consist of a $q$-ary bit deposited at each point of the
boundary $\P^1(\K)$. We want to transform this classical code built using the
algebro-geometric properties of the Bruhat--Tits tree, into a quantum error correcting
code that generates a holographic code for the Bruhat--Tits tree and its boundary
at infinity. 

\smallskip
\subsection{Classical Algebro-Geometric Codes for Mumford Curves}

The construction above can be generalized in the case of Mumford curves.
Let $\Gamma$ be a $p$-adic Schottky group and 
$\Omega_\Gamma=\P^1(\K)\smallsetminus \Lambda_\Gamma$
the domain of discontinuity of $\Gamma$ acting on the boundary $\P^1(\K)$, the complement of the limit set
$\Lambda_\Gamma$. The quotient $X=\Omega_\Gamma/\Gamma$ is a Mumford curve of genus $g$ equal
to the number of generators of the Schottky group. Unlike complex Riemann surfaces, which always
admit a Schottky uniformization, only very special $p$-adic curves admit a Mumford curve uniformization.
Indeed, these curves must have the property that their reduction mod $\m$ is totally split: as a curve over $\F_q$
it consists of a collection of $\P^1$'s with incidence relations described by the dual graph $G$. This is the
finite graph at the center of the quotient $\cT_\K /\Gamma$, obtained as the quotient 
$G=\cT_\Gamma/\Gamma$,
where $\cT_\Gamma$ is the subtree if $\cT_\K$ spanned by the geodesic axes of the hyperbolic elements
$\gamma\neq 1$ of $\Gamma$, with $\partial\cT_\Gamma =\Lambda_\Gamma$.

\smallskip

Using the identification between the finite graph $G$ and the dual graph of the reduction curve,
we can again associate to each vertex in $G$ a copy of $\P^1(\F_q)$ (the corresponding component
in the curve), and to each of these projective lines a Reed--Solomon code as in the previous
construction. Now, however, we need to impose compatibility conditions between these codes
at the incidence points between different components of the curve, that is, along the edges
of the finite graph $G$.  Thus, we associate to the finite graph $G$ a classical code $C(G)$
constructed as follows. Start with a code $\hat C_{w,k}$ associated to each vertex $v$, which
inputs $a=(a_0,\ldots,a_{k-1})$ and outputs $f_a(u,v)=\sum_i a_i  u^i v^{k-1-i}$ at each point
$x=(u:v)$ of the associated $\P^1(\F_q)$.
Consider the set $\cF$ of functions $f=(f_1,\ldots, f_N)$, with $N=\# V(G)$, and $f_i$ a homogeneous
polynomial of degree $\deg(f_i)<k$ on the $i$-th component $\P^1(\F_q)$, with the property that if
$x=(u_i:v_i)=(u_j:v_j)$ is an intersection point between the $i$-th and the $j$-th components of the
reduction curve, then $f_i(u_i:v_i)=f_j(u_j:v_j)$. Thus, each edges $e\in E(G)$ imposes a relation
between $f_i$ and $f_j$ which requires the value that the codes $\hat C_{w,k}$ at the vertices 
$\nu_i$ and $\nu_j$
deposit at the point $x$ to be the same. Thus, the resulting code $C(G)$ is an $\F_q$-linear code
with input $\F_q^{kN -M}$ where $N=\# V(G)$ and $M=\# E(G)$. We have $kN-M=(k-1)N +1-b_1(G)$
hence we need to assume $k>1+(b_1(G)-1)/N$.

\smallskip

The free legs of the graph $G$ are all the legs that point towards the infinite trees in
$\cT_\K/\Gamma$ that extend from the vertices of $G$ to the boundary Mumford
curve $X(\K)=\partial \cT_\K/\Gamma$. At each vertex along these trees we consider
Reed-Solomon codes as in the case of $\P^1(\K)$, with one input coming from the
previous vertex closer to $G$ and $k-1$ new inputs and outputs at the $q$ forward 
pointing legs.  This determines a classical code associated to the infinite graph 
$\cT_\K/\Gamma$, with logical inputs at the vertices and outputs at the points of the
Mumford curve $X(\K)$. The finite graph $G$ and the infinite graph $\cT_\K/\Gamma$
containing it are a genus $g$ generalization of the $p$-adic BTZ black hole, which
corresponds to the $g=1$ case of Mumford--Tate elliptic curves.

\smallskip
\subsection{Reed--Solomon Codes and Quantum Algebro-Geometric Codes}

There is a general procedure for passing from classical codes to quantum codes,
based on the Calderbank--Rains--Shor--Sloane algorithm \cite{CRSS}, see also
\cite{AsKn}. It can be applied to certain classes of algebro-geometric codes and 
in particular to generalized Reed--Solomon codes. 

\smallskip

Let $\cH=\C^q$ be the Hilbert space of a single $q$-ary qubit and
$\cH_n=(\C^q)^{\otimes n}$ the space of $n$ $q$-ary qubits. We label
an orthonormal basis of $\cH$ by $|a\rangle$ with $a\in \F_q$. Thus,
a $q$-ary qubit is a vector $\psi =\sum_{a\in \F_q} \lambda_a \,|a\rangle$ with
$\lambda_a\in \C$, and an $n$-tuple of $q$-ary qubits is given by a vector
$\psi=\sum_{a=(a_1\ldots a_n) \in \F_q^n} \lambda_a |a\rangle$ where
$|a\rangle =|a_1\rangle \otimes \cdots \otimes |a_n\rangle$.
Quantum error correcting codes are subspaces $\cC$ of $\cH_n$ that are
error correcting for a certain number of ``q-ary bit flip" and ``phase flip" errors.
More precisely, an error operator $E$ is detectable by a quantum code $\cC$
if $P_\cC E P_\cC =\lambda_E \, P_\cC$, where $P_{\cC}$ is the orthogonal
projection onto the code subspace and $\lambda_E$ is a scalar. In particular,
one considers error operators that affect up to a certain number of qubits
in an $n$-qubits state, namely error operators of the form $E=E_1\otimes \cdots \otimes E_n$,
of weight $\omega(E)=\# \{ i\,:\, E_i\neq I \}$. The minimum distance $d_Q(\cC)$
of the quantum code is the largest $d$ such that all errors with $\omega(E)< d$
are detectable. 

\smallskip

The bit and phase flip error operators are defined on a single $q$-ary qubit as
$$ T_b |a\rangle = |a+b \rangle, \ \ \  R_b |a\rangle = \xi^{\Tr(\langle a,b \rangle)}|a\rangle, $$
where $\xi$ is a $p$-th primitive root of unity, and 
$\Tr: \F_q \to \F_p$ is the trace function, $\Tr(a)=\sum_{i=0}^{r-1} a^{p^i}$, with 
$\langle a,b \rangle = \sum_{i=1}^{r} a_i b_i$ and with $R^{b_i} |a_j\rangle = \xi^{\Tr(a_j b_i)}|a_j\rangle$.
Let $\{ \gamma_i \}_{i=1}^r$ be
a basis of $\F_q$ as an $\F_p$-vector space, so that $a=\sum_i a_i \gamma_i$
and $b=\sum_i b_i \gamma_i$. Then the error operators $T_b$ and $R_b$
can be written respectively as
$$ T_b =T^{b_1}\otimes \cdots \otimes T^{b_r}, \ \ \ R_b= R^{b_1}\otimes \cdots \otimes R^{b_r} $$
with $T$ and $R$ given by the operators acting on $\C^p$ of matrix form
$$ T=\begin{pmatrix} 0 & 1 & 0 & \cdots & 0 \\ 0 & 0 & 1 & \cdots & 0 \\ \vdots & & & \ddots & \vdots \\
0 & 0 & 0 & \cdots & 1 \\
1 & 0 & 0 & \cdots & 0  \end{pmatrix} \ \ \  R = \begin{pmatrix} 
1 &  & & & \\     & \xi & & & \\   &  & \xi^2 & & \\   &  &  & \ddots & \\   & & & & \xi^{p-1} 
\end{pmatrix}$$
satisfying the commutation relation $TR =\xi RT$. The operators $T_aR_b$ with $a,b\in \F_q$
form an orthonormal basis for $M_{q\times q}(\C)$ under the inner product 
$\langle A, B\rangle=q^{-1} \Tr(A^* B)$, hence these operators generate all possible quantum
errors on the space $\C^q$ of a single $q$-ary qubit. The action of error operators on a 
state of $n$ $q$-ary qubits can similarly be written in terms of operators $T_a R_b$ with
$$  E_{a,b}=T_a R_b = (T_{a_1} \otimes \cdots \otimes T_{a_n}) (R_{b_1}\otimes \cdots \otimes R_{b_n}), $$
for $a=(a_1,\ldots, a_n), b=(b_1\ldots,b_n)\in \F_q^n$. The operators $E_{a,b}$ satisfy $E^p_{a,b}=I$ and
the commutation and composition rules
$$ E_{a,b} E_{a',b'} = \xi^{\langle a,b'\rangle - \langle b, a' \rangle} E_{a',b'} E_{a,b}, \ \ \ 
E_{a,b} E_{a',b'} = \xi^{-\langle b,a'\rangle} E_{a+a',b+b'}, $$
where $\langle a,b \rangle=\sum_i \langle a_i, b_i\rangle=\sum_{i,j} a_{i,j} b_{i,j}$, 
with $a_i,b_i\in \F_q$, written as $a_i=\sum_j a_{i,j} \gamma_j$
and $b_i=\sum_j b_{i,j} \gamma_j$, after identifying
$F_q$ as a vector space with $\F_p^r$.
Thus, we can consider the group $\cG_n = \{ \xi^i E_{a,b}, \, a,b\in \F_q^n, \, 0\leq i \leq p-1 \}$ of
order $pq^{2n}$.  A quantum stabilizer error-correcting code $\cC$ is a subspace $\cC\subset \cH_n$ 
that is a joint eigenspace of operators $E_{a,b}$ in an abelian subgroup $\cS\subset \cG_n$. 

\smallskip

Let $\varphi\in \Aut_{\F_p}(\F_p^r)$ be an automorphism. In particular, we consider $\varphi$ given by
the trace as in \cite{AsKn}, so that we the associated pairing is 
$$ \langle (a,b), (a',b') \rangle =\langle a, \varphi(b')\rangle - \langle a', \varphi(b)\rangle =
 \Tr(\langle a,b'\rangle_* - \langle a',b\rangle_*), $$
 where, for $a, b \in \F_q^n$, the inner product $\langle a,b \rangle \neq \langle a,b \rangle_*$, since $\langle a,b \rangle_* = \sum_{i=1}^n a_i b_i$, while $\langle a,b \rangle = \sum_{i=1}^n \langle a_i, b_i \rangle = \sum_{i=1}^n \sum_{j=1}^r a_{i,j} b_{i,j}$.
 If $C\subset \F_q^{2n}$ is a classical self-orthogonal code with respect to this pairing,
then the subgroup $\cS \subset \cG_n$ given by the elements $\xi^i E_{a,\varphi(b)}$ with $(a,b)\in C$
is an abelian subgroup of $\cG_n$, because of the commutation rule above. This construction is the
CRSS algorithm that associates to a self-orthogonal classical $[2n,k,d]_q$ code a stabilizer quantum
$[[n,n-k,d_Q]]_q$-code, where $d_Q=\min\{ \omega(a,b)\,:\, (a,b)\in C^\perp \smallsetminus C \}$,
where the weight $\omega(a,b)=\#\{ i\,:\, a_i\neq 0 \text{ or } b_i\neq 0 \}$, and $C^\perp=\{ (v,w)\in \F_q^{2n}\,:\, \langle (a,b),(v,w)\rangle =0, \, \forall (a,b)\in C\}$.

\smallskip

We can view the CRSS algorithm assigning the quantum stabilizer code $\cC$ to the classical code $C$
as an encoding process that takes the $q^k$ input vectors $(v,w) \in \F_q^{2n}$ of the classical code $C$
and encodes the states $|(v,w) \rangle$ using the vectors $\psi  \in (\C^q)^{\otimes n}$ satisfying 
$E_{v,\varphi(w)}\psi =\lambda \psi$ in a common eigenspace of the $E_{v,\varphi(w)}$.

\smallskip

A slightly more general version of the CRSS algorithm starts with two classical linear 
$q$-ary codes $C_1 \subseteq C_2$ of length $n$ and dimensions $k_1$ and $k_2$ and
associates to them a quantum code $\cC=\cC(C_1,C_2)$ with parameters 
$[[n, k_2-k_1, \min\{ d(C_2\smallsetminus C_1), d(C_1^\perp\smallsetminus C_2) \}]]_q$,
see \cite{Gra2}, \cite{KimWal}. The procedure for the construction of the quantum code
is similar to the version of the CRSS algorithm recalled above. One constructs a code
$C=\gamma C_1 + \bar \gamma C_2$ in $\F_{q^2}^n$ with $\gamma$ a primitive element
of $\F_{q^2}$ and $\{ \gamma, \bar\gamma \}$ a linear basis of $\F_{q^2}$ as an $\F_q$-vector
space. By identifying $\F_{q^2}^n$ as an $\F_q$-vector space with $\F_q^{2n}$, we obtain
a self orthogonal $C \subset \F^{2n}_q$, to which the CRSS algorithm discussed before can
be applied.

\smallskip

Conditions under which Reed-Solomon codes satisfy a self-dual condition, and the
corresponding quantum Reed-Solomon codes obtained via a CRSS type algorithm
are analyzed, for instance, in \cite{Gra} and \cite{Gua}.
We use here a construction of \cite{AsKn}, which shows that, if $C$ is a $q^2$-ary classical 
$[n,k,d]_{q^2}$-code, which is Hermitian-self-dual, then there exists an associated $q$-ary
$[[n,n-2k,d_Q]]_q$-quantum code, with $d_Q\geq d$. Here Hermitian-self-dual means that
the classical code $C$ is self dual with respect to the ``Hermitian" pairing 
$$ \langle v,w \rangle_H = \sum_{i=1}^n v_i w_i^q, \ \ \text{ for } v,w \in \F_{q^2}^n. $$
This is a variant of the CRSS algorithm described above, where a Hermitian-self-dual
code of length $n$ over the field extension $\F_{q^2}$ is used to construct a self-dual code 
$\tilde C$ of length $2n$ over $\F_q$ to which the CRSS algorithm can be applied,
obtained by expanding the code words $v\in C$ using a basis $\{ 1,\gamma \}$ of $\F_{q^2}$
as a $\F_q$-vector space, where $\gamma$ is an element in $\F_{q^2}\smallsetminus \F_q$
satisfying $\gamma^q =-\gamma +\gamma_0$ for some fixed $\gamma_0\in \F_q$. Using this
approach, it suffices to construct generalized Reed--Solomon codes $\hat C_{w,k}$ of length $n<q+1$
over $\F_{q^2}$ that are Hermitian self-dual, in order to obtain associated quantum codes 
$\hat \cC_{w,2n-k}$ as a code subspace of the $n$ $q$-ary qubits space $\cH_n=(\C^q)^{\otimes n}$. 
It is possible to ensure the Hermitian self-duality condition for generalized Reed--Solomon codes
by taking the weights vector $w=(w_1,\ldots,w_n)\in (\F_{q^2}^*)^n$ to satisfy 
$\sum_{i=1}^n w_i^{q+1} x_i^{qj+\ell}=0$ for all $0\leq j,\ell\leq k-1$, where $x=(x_1,\ldots,x_n) \in \F_{q^2}^n$
are the $n$ chosen points (excluding $\infty$) of $\P^1(\F_{q^2})$.  Using this method, it is
proved in \cite{LXW} that the choice $w_i=1$, with $n=q^2=\# \F_{q^2}$ and $k=q$, produces a 
Reed-Solomon code $C=C_{1,q}$ that is Hermitian-self-dual, and an associated 
$[[q^2+1, q^2-2q+1,q+1]]_q$-quantum Reed-Solomon code $\hat \cC$. Moreover, it is also shown in \cite{LXW}
that for $w_i$ satisfying $w_i^{q+1}=(\prod_{j\neq i} (x_i - x_j))^{-1}$, with $n\leq q$ and 
$x=(x_1,\ldots,x_n) \in \F_q$, and for $k\leq \lfloor n/2 \rfloor$, the generalized Reed-Solomon codes satisfy 
$C_{w,k}\subseteq C_{w,n-k}$ and $C_{w,k}$ is hermitian self dual to $C_{w,n-k}$. Thus,  hence the CRSS algorithm 
can be applied to obtain an $[[n,n-2k,k+1]]_q$-quantum Reed-Solomon code 
$\cC_{w,k}=\cC(C_{w,k},C_{w,n-k})$.

\smallskip
\subsection{The case of the perfect tensors}

In particular, from the construction described above we see that we obtain the case
of perfect tensors as the special case where $n=q$ and $k=(q-1)/2$. We obtain this
using the generalized Reed--Solomon codes as in Theorem~6 of \cite{LXW}, for the
case $n\leq q$ and $k\leq \lfloor \frac{n}{2} \rfloor$, with a choice 
of the weights $w_i$ satisfying $w_i^{q+1}=(\prod_{j\neq i} (x_i - x_j))^{-1}$, with $n=q$ and 
$x_i \in \F_q$.  As shown in Theorem~6 of \cite{LXW}, 
this produces two classical generalized Reed-Solomon codes $C_{w,\frac{q-1}{2}}\subseteq C_{w,\frac{q+1}{2}}$
that are hermitian self-dual. The associated quantum generalized Reed--Solomon code is then obtained
via the general construction of Ashikhmin--Knill (Theorem~4 and Corollary~1 of \cite{AsKn})
that associates to a classical $[n,k,d]_{q^2}$ code contained in its hermitian dual a quantum $[[n,n-2k,d]]_q$ code. One can see directly that, in the case of perfect tensors
when $n=q$, the weights are constant and given by $w_i^{q+1} = p-1$ for all $i=1, \ldots, q$. 

\smallskip

Thus, we can regard the construction described above with generalized Reed--Solomon codes
as a generalization of the usual construction of perfect tensors, which recovers the perfect tensor
case for a particular choice of (constant) weights of the classical Reed--Solomon codes. 

\smallskip

The more general cases with non-constant weights assign different weights to different directions
in the Bruhat--Tits tree. These may be useful in view of holographic models where the bulk
geometry is dynamical, as in \cite{GHJMPSST}, and also described by different weights 
in different directions in the tree.

\smallskip
\subsection{Holographic quantum codes on Bruhat--Tits trees and
Mumford curves}

We use the procedure described above to pass from classical algebro-geometric codes,
in particular generalized Reed-Solomon codes, to associated quantum stabilizer codes,
to construct a holographic code on the Bruhat--Tits tree $\cT_\K$ associated to the
classical codes constructed above. 

\smallskip

We have seen above that, in order to apply the CRSS algorithm, we pass to a
quadratic extension $\F_{q^2}$ of $\F_q$ and consider Reed-Solomon codes over
$\F_{q^2}$. In terms of the Bruhat--Tits tree, we can pass to an unramified quadratic 
extension $\bL$ of the field $\K$, so that the Bruhat--Tits tree $\cT_\bL$ is obtained
from the Bruhat--Tits tree $\cT_\K$ simply by adding new branches at each vertex,
so as to obtain a homogeneous tree of valence $q^2+1$. Since the extension
is unramified, it is not necessary to insert new vertices along the edges, and we can view the
tree $\cT_\bK$ as a subtree of $\cT_\bL$. We then proceed
to construct a classical code associated to $\cT_\bL$ using Reed-Solomon codes 
$\hat C_{w,k}$ placed at the vertices according to the procedure described in the 
previous sections. Using the construction above, with $w_i=1$ and $k=q$, we associate to each vertex a
quantum Reed-Solomon code $\hat\cC$ with code parameters $[[q^2+1, q^2-2q+1,q+1]]_q$.
This corresponds to considering a state space $\cH_{q^2+1}=(\C^q)^{\otimes q^2+1}$ associated to
each vertex, which we can think of as a state with a $q$-ary qubit sitting at each of the
$q^2+1$ points of $\P^1(\F_{q^2})$, or equivalently at each of the legs surrounding that 
vertex in $\cT_\bL$. The quantum code $\hat\cC$ detects quantum errors of weight
up to $q+1=\# \P^1(\F_q)$. Thus, by identifying $\cT_\K\subset \cT_\bL$ and $\P^1(\F_q)\subset \P^1(\F_{q^2})$
as the set of directions along the subtree $\cT_\K$, we can arrange that the code $\hat\cC$ corrects 
quantum errors along the $\cT_\K$ directions. One can also use a bipartition $A\cup A^c$ of the edges at each
vertex of $\cT_\bL$, with $\# A =k$ and associate to the bipartition a pair of codes
$\hat C_{w,k}$ and $\hat C_{w,n-k}$ with associated quantum Reed-Solomon codes 
$\hat C_{w,k}$ as above at the vertices of $\cT_\K$.

\smallskip

One can think of the classical codes $\hat C$ associated to the vertices of the Bruhat--Tits 
tree in this way as performing an encoding of $q+1$ classical $q^2$-ary bits associated to the points of
$\P^1(\K)$ into $q^2+1$ classical $q^2$-ary bits associated to the points of $\P^1(\bL)$. Thus,
the whole classical code associated to this quadratic extension can be seen as a way of encoding 
a state consisting of classical $q^2$-ary bits associated to the edges of $\cT_\K$ into
a set of classical $q^2$-ary bits associated to the edges of $\cT_\bL$, and the
letter into a state of $q^2$-ary bits associated to the set of boundary points $\P^1(\bL)$. 
The corresponding CRSS quantum codes $\hat\cC$ at the vertices of the Bruhat--Tits tree $\cT_\bL$
encode the input given by the common eigenspace of the error operators associated to the
code words of the classical code $\hat C$ into a state consisting of a $q$-ary qubit
placed at each leg around the vertex.

\smallskip

In order to combine these quantum codes placed at the vertices of the Bruhat--Tits tree
$\cT_\bL$ into a holographic code over the whole tree, with logical inputs in the bulk
and physical outputs at the boundary $\P^1(\bL)$, notice that at each vertex $v$ we have
the same subspace $\cH_\nu$ given by the common eigenspace 
$\cH_\nu=\{ \psi \,:\, E_{v,\varphi(w)}\psi =\lambda \psi \}$ for all words $(v,w)$ in the
classical code. We encode states $\psi_\nu \in \cH_\nu$ as $\psi_\nu =(\psi_{\nu,x})_{x\in \P^1(\F_{q^2})}$,
where the points $x\in \P^1(\F_{q^2})$ label the legs around the vertex $\nu$, so that we think of
$\psi_{\nu,x}\in \C^q$ as the $q$-ary qubit deposited on the leg $x$ by the quantum code $\hat\cC$ 
sitting at the vertex $\nu$. Starting at the root vertex and proceeding towards the outside of the tree,
at each next step, the leg $\infty \in  \P^1(\F_{q^2})$ around the new vertex $\nu$ is the one connected to
a leg $x_i \in \F_{q^2} \subset \P^1(\F_{q^2})$ of the previous vertex $\nu'$, which receives an output 
$\psi_{\nu',i}$. Thus, the $q$-ary qubit $\psi_{\nu,\infty}$ is determined as it has to match the
output $\psi_{\nu',i}$ of the previous code, while the remaining possible inputs correspond to
the choices of $\psi \in \cH_v$ with that fixed $\psi_{v,\infty}$ component. Proceeding towards
the boundary of the tree determines a holographic code on $\cT_\bL$ that outputs $q$-ary
qubits at the points of $\P^1(\bL)$. As mentioned above, the quantum code detects errors 
along the subtree $\cT_\K$. As in the case of the classical codes, there is an asymmetry
in this construction of the holographic code between the roles of the root vertex and of the 
remaining vertices of the Bruhat--Tits tree.

\section{Discrete and continuous bulk spaces: Bruhat-Tits buildings and Drinfeld symmetric spaces}

Unlike its Archimedean counterparts, either Euclidean AdS$_2$/CFT$_1$ with bulk $\H^2$
and boundary $\P^1(\R)$ or Euclidean AdS$_3$/CFT$_2$ with bulk $\H^3$ 
and boundary $\P^1(\C)$, the $p$-adic AdS/CFT correspondence has two different choices of
bulk spaces (one discrete and one continuous) which share the same conformal boundary at infinity. The discrete
version of the bulk space is given by the Bruhat--Tits tree $\cT_\K$ of $\PGL(2,\K)$, with $\K$ a finite extension
of $\Q_p$, while the continuous form of the bulk space is given by Drinfeld's $p$-adic upper half plane $\Omega$.
Both have the same boundary $\P^1(\K)$.  We argue here that the full picture of the
$p$-adic AdS/CFT correspondence should take into account both of these bulk spaces and the
relation between them induced by the norm map.

\smallskip

The rank-two case can be generalized to higher rank, with the Bruhat--Tits buildings of $\PGL(n,\K)$
generalizing the Bruhat--Tits tree and the higher dimensional Drinfeld symmetric spaces 
generalizing the Drinfeld upper half plane, see \cite{Kato}. 

\subsection*{The geometry of the Drinfeld plane}

We review quickly the geometry of the Drinfeld upper half plane, see \cite{BouCar}.
We denote by $\K$ a finite extension of $\Q_p$ and by $\C_p$ the completion of the
algebraic closure of $\K$. Drinfeld's $p$-adic upper half plane is the space
$$ \Omega = \P^1(\C_p) \smallsetminus \P^1(\K). $$
We also denote by $\cT_\K$ the Bruhat--Tits tree of $\PGL(2,\K)$, with boundary at infinity $\P^1(\K)$.
It is convenient to think of $\P^1(\C_p)$ as the set of classes, up to homotheties in $\C_p^*$,
of non-zero $\K$-linear maps $\varphi: \K^2 \to \C_p$, with $\P^1(\K)$ the set of classes of maps as
above with $\K$-rank equal to one. This can be seen by identifying points $(\alpha:\beta)$ of $\P^1$ with
homogeneous ideals $\langle y\alpha - x \beta \rangle$ in the polynomial ring in the variables $(x,y)$.
The $\K$-linear map $\varphi: \K^2 \to \C_p$ given by $\varphi(x,y)=y\alpha - x \beta$ has a
non-trivial kernel when $\alpha/\beta \in \K$ (assuming $\beta\neq 0$) 
and is invertible if $\alpha/\beta \in \C_p\smallsetminus \K$. 
Thus, $\P^1(\C_p) \smallsetminus \P^1(\K)$ can be identified with the set of homothety classes 
of invertible $\K$-linear maps $\varphi: \K^2 \to \C_p$.

\smallskip

Given such an injective linear map, one can then compose it with the norm on $\C_p$. Recall
that the Bruhat--Tits tree can be defined in terms of equivalence classes of norms. Namely,
vertices of the Bruhat--Tits tree correspond to classes of lattices $M$ in $\K^2$ up to similarity, namely
$M_1\sim M_2$ if $M_1=\lambda M_2$ for some $\lambda\in \K^*$. To a lattice $M$ one
associates a norm $|\cdot |_M$, namely a real valued function on $\K^2$ which is positive on
non-zero elements, satisfies $| a\cdot \underline{x} |_M =|a|\cdot | \underline{x} |_M$ for all $a\in \K$ and 
$\underline{x} \in \K^2$, with $|a|$ the $p$-adic norm on $\K$, and $| \underline{x}+\underline{y} |_M\leq
\max\{ | \underline{x}|_M, |\underline{y} |_M \}$. The norm $|\underline{x}|_M$ is defined as follows. 
Let $\pi$ be a uniformizer in $\cO_\K$ such that $k=\cO_\K /\pi\cO_\K$ is the residue field $k=\F_q$.
The fractional ideal $\{ \lambda \in \K\,:\, \lambda \underline{x} \in M \}$ is generated by a power $\pi^m$.
The norm is then defined as $| \underline{x}|_M=q^m$ on non-zero vectors. Equivalent norms
$|\cdot |_{M_1} =\gamma |\cdot |_{M_2}$ for $\gamma\in \R^*_+$ correspond to equivalent lattices. 
Two vertices in the Bruhat--Tits tree are adjacent iff the corresponding equivalence classes
of lattices have representatives satisfying $\pi M \subset M' \subset M$. To see this in terms
of norms, we can choose an $\cO_\K$-basis $\{ e_1, e_2 \}$ for $M$ and $\{ e_1 , \pi e_2 \}$ for $M'$.
Then $| x e_1 + y e_2 |_M = \max \{ |x|, |y| \}$ and $| x e_1 + y e_2 |_{M'} =\max \{ |x|, |\pi|^{-1} \cdot |y| \}$.
The edge $e$ between the vertices $v=[M]$ and $v'=[M']$ is then parameterized by the classes of
norms $| x e_1 + y e_2 |_t = \max \{ |x|, |\pi|^{-t} \cdot |y| \}$ for $0\leq t \leq 1$, see \cite{Kato}.
This description of the Bruhat--Tits tree in terms of equivalence classes of norms on $\K^2$ determines
a map from the Drinfeld uppar half plane to the Bruhat--Tits tree, directly induced by the norm.
Namely, given a point in $\Omega$, which we identify as above with an invertible $\K$-linear
map $\varphi: \K^2 \to \C_p$, we obtain a surjective map 
$$ \Upsilon: \Omega \to \cT_\K $$
by setting $\Upsilon(\varphi)=|\cdot |_\varphi$, where $|\cdot |_\varphi$ is the norm
on $\K^2$ defined by $|\underline{x} |_\varphi = |\varphi(\underline{x})|$, where the norm
on the right-hand-side if the $p$-adic norm on $\C_p$. The explicit form of this map is
discussed in \cite{BouCar}. We identify a point $(\zeta_0:\zeta_1) \in \P^1(\C_p)\smallsetminus \P^1(\K)$ 
with the map $\varphi: \K^2 \to \C_p$ that
maps $x e_1 + y e_2 \mapsto x \zeta_0 + y \zeta_1 \in \P^1(\C_p)$. In an affine patch (say with $\zeta_1\neq 0$)
we can write the homotethy class of $\varphi$ as $x e_1 + y e_2 \mapsto x \zeta + y \in \C_p\smallsetminus \K$.
Then the preimages under the map $\Upsilon$ of two adjacent vertices $v,v'$ of $\cT_\K$
and the edge $e$ connecting them are given, respectively, by 
$$ \Upsilon^{-1}(v) = \{ \zeta \in \C_p \,:\, |\zeta|\leq 1\} \smallsetminus \bigcup_{a\in \cO_\K/\pi \cO_\K} \{ \zeta \in \C_p \,:\,  
|\zeta - a| < 1 \} $$
$$ \Upsilon^{-1}(v') =\{ \zeta \in \C_p \,:\, |\zeta|\leq q^{-1} \} \smallsetminus \bigcup_{b \in \pi \cO_\K/\pi^2 \cO_\K} 
\{ \zeta \in \C_p \,:\,   |\zeta - b| < q^{-1} \} $$
where $v=[M]$, $v'=[M']$ with $\pi M \subset M' \subset M$, and for $e_t = (1-t) v + t v'$, for $0<t<1$, along the edge $e$
$$ \Upsilon^{-1}(e_t) =\{ \zeta \in \C_p \,:\, |\zeta|\leq q^{-t} \}, $$
while
$$ \Upsilon^{-1}(e) = \{ \zeta \in \C_p \,:\, |\zeta|\leq 1\} \smallsetminus \bigcup_{a\in (\cO_\K \smallsetminus \pi \cO_\K)/\pi \cO_\K} \{ \zeta \in \C_p \,:\,  |\zeta - a| < 1 \} $$ $$ \smallsetminus \bigcup_{b \in \pi \cO_\K/\pi^2 \cO_\K} 
\{ \zeta \in \C_p \,:\,   |\zeta - b| < q^{-1} \}. $$
For a detailed proof of this fact we refer to \S 2 of \cite{BouCar}. A part of the Drinfeld plane
corresponding to the regions $\Upsilon^{-1}(v)$, $\Upsilon^{-1}(v')$ and $\Upsilon^{-1}(e)$
with $\partial e=\{ v,v'\}$ in the Bruhat--Tits tree can be illustrated as follows (from \cite{BouCar}):
\begin{center}
\includegraphics[scale=0.5]{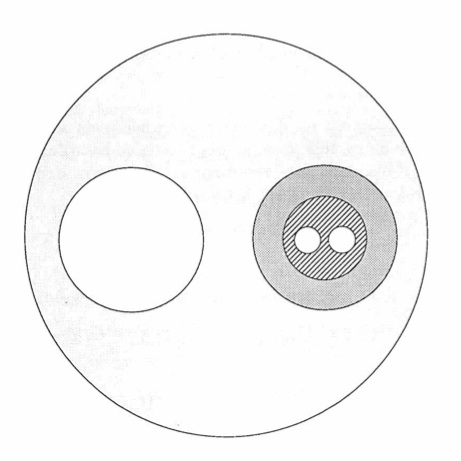} 
\end{center}  
where the light colored region is  $\Upsilon^{-1}(v)$, the striped shaded region is $\Upsilon^{-1}(v')$,
and the dark shaded cylinder connecting them is $\Upsilon^{-1}(e)$. Thus, one can visualize the
Drinfeld plane as a continuum that is a ``tubular neighborhood" of the discrete Bruhat--Tits tree,
with the regions $\Upsilon^{-1}(v)$ viewed as the $p$-adic analog of pair-of-pants
decompositions for complex Riemann surfaces. A lift of the projection map $\Upsilon$ to the
Bruhat-Tits tree realizes the tree as a skeleton of the Drinfeld plane.

\subsection*{Higher rank buildings and Drinfeld symmetric spaces}

An analogous description holds relating the Bruhat--Tits buildings $\cT_{n,\K}$ of $\PGL_{n+1}(\K)$,
with $\K$ a finite extension of $\Q_p$ and the associated Drinfeld symmetric space
$$ \Omega_n = \P^n(\C_p)\smallsetminus \cup_{H \in \cH_\K} H, $$
where $\cH_\K$ is the set of all $\K$-rational hyperplanes in $\P^n(\C_p)$. There is again
a map $\Upsilon_n : \Omega_n \to \cT_{n,\K}$ where the preimages of simplices in the
Bruhat-Tits building is described in terms of norm conditions, \cite{Kato}.

\smallskip

The Bruhat--Tits building $\cT_{n,\K}$ of $\PGL_{n+1}(\K)$ is a simplicial complex with vertex set
$V(\cT_{n,\K})=\cT_{n,\K}^0$ given by the similarity classes $M_1\sim M_2$ if $M_1=\lambda M_2$ 
for $\lambda\in \K^*$ of lattices in an $n+1$ dimensional vector space $V$ over $\K$. A set
$\{ [M_0], \ldots, [M_\ell] \}$ of such classes defines an $\ell$-simplex in $\cT_{n,\K}^\ell$ in the
Bruhat--Tits building iff $M_0\supsetneq M_1 \supsetneq M_2 \supsetneq \cdots \supsetneq M_\ell \supsetneq \pi M_0$,
with $\pi \in \cO_\K$ a prime element with $\F_q=\cO_\K/\pi \cO_\K$ the residue field. Such a sequence
determines a flag $\bar M_0 \supsetneq \bar M_1 \supsetneq \cdots \supsetneq \bar M_\ell \supseteq 0$
of subspaces $\bar M_i =M_i/\pi M_i$ of an $n+1$-dimensional $\F_q$-vector space. The $\ell$-simplices
in $\cT_{n,\K}^\ell$ containing a given vertex $[M]$ are in one-to-one correspondence with such 
flags with $[M_0]=[M]$. As before, we consider norms on $V\simeq \K^{n+1}$ and similarity classes of
norms. There is a $\PGL_{n+1}(\K)$-equivariant homeomorphism between the resulting space of
equivalence classes of norms and the geometric realization of the simplicial complex $\cT_{n,\K}$.

\smallskip

Consider then points $\zeta=(\zeta_0:\cdots : \zeta_n)\in \P^n(\C_p)$ and the map $\varphi: V\to \C_p$
given by $\sum_{i=0}^n a_i e_i \mapsto \sum_{i=0}^n a_i \zeta_i$. The map $| \sum_{i=0}^n a_i e_i |_\varphi
=| \sum_{i=0}^n a_i \zeta_i |$ determines an equivalence class of norms iff the point $\zeta \in \P^1(\C_p)$
does not lie in any $\K$-rational hyperplane. This determines the map $\Upsilon: \Omega_n \to \cT_{n,\K}$
that generalizes in higher rank the map from the Drinfeld plane to the Bruhat-Tits tree. As in the previous case,
one can describe the preimages under this map. For example, the preimage of a vertex $v=[M]$ is
given by 
$$ \Upsilon^{-1}(v)=\{ |\zeta_0|=\cdots =|\zeta_n|=1\} \smallsetminus \cup_H \{ \zeta \mod \pi\in H \} $$
with the union over hyperplanes and 
$$ \Upsilon^{-1}(e_t)=\{ |\zeta_0|=\cdots =|\zeta_{n-1}|=1, \, |\zeta_n|=q^{-t} \} $$
for $e_t$ point along an edge $e$, with $0<t<1$, see \S 2 of \cite{Kato} for more details.

\section{Tensor networks on the Drinfeld plane}\label{DplaneSec}

Because the $p$-adic AdS/CFT correspondence has two different choices of bulk space,
in addition to considering classical and quantum codes associated to the Bruhat--Tits tree
in constructing a version of tensor networks, we can also work with the Drinfeld $p$-adic
upper half plane. Because this is a continuous rather than a discrete space, the type of
construction we can consider there will be more similar to the type of construction of
tensor networks on the ordinary upper half plane (the $2$-dimensional real hyperbolic
plane $\H^2$) described in \cite{HaPPY}. The map $\Upsilon$ from the Drinfeld upper
half plane to the Bruhat--Tits tree will then make it possible to relate the construction of
tensor networks on the first to the latter. To this purpose, we start by reviewing the construction
of the pentagon holographic code from \cite{HaPPY}.

\smallskip
\subsection{Pentagon Code on the Real Hyperbolic Plane}

In \cite{HaPPY} a holographic code is constructed using a tessellation of the
real hyperbolic plane $\H^2$ by pentagons, with quantum codes given by a
six leg perfect tensor placed at each tile. Unlike the codes discussed in the
previous section on Bruhat--Tits trees, this code has no preferred base point
in the tiling and all tiles are treated equally, and the codes are symmetric
with respect to permutations of the five legs places across the edges of
the tiles, thus preserving the full symmetry group of the tiling. We discuss
briefly some aspects of this pentagon code here before turning to analogous
constructions on the Drinfeld $p$-adic upper half plane. 

\smallskip 

The real hyperbolic plane $\H^2$ (which we can conveniently represent as
the Poincar\'e disk) has a regular periodic tessellation by right-angle pentagons.
\begin{center}
\includegraphics[scale=0.35]{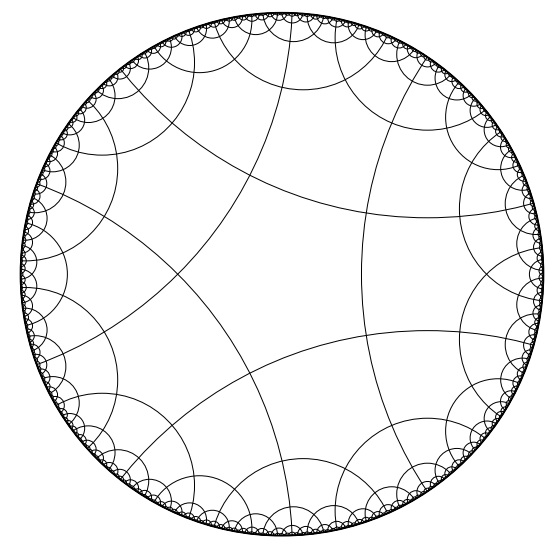} 
\end{center}
The corresponding symmetry group is the Fuchsian group $\Gamma \subset \PSL(2,\R)$
of signature $(2,2,2,2,2)$ generated by the reflections about the sides of a single right-angled 
hyperbolic pentagon. An interesting property of this Fuchsian group, from the algebro-geometric
perspective is the fact that, if one subdivides an equilateral right-angled hyperbolic pentagon
into $10$ triangles with angles $\pi/2, \pi/4, \pi/5$, then one can realize the group $\Gamma$
as a finite index subgroup of a triangle Fuchsian group $\Gamma'$ of signature $(2,4,5)$. These 
Fuchsian groups have the property that the quotient Riemann surfaces $\H/\Gamma=X$ is
arithmetic as an algebraic curve (that is, it is defined over a number field), \cite{Cohen}. 

\smallskip

The construction of the pentagon holographic code in \cite{HaPPY} places over each tile
of this right-hangled pentagon tiling a quantum code given by the six leg perfect tensor
determined by a $5$-qubit $[[5,1,3]]_2$-quantum code
$$ \cC \subset \cH^{\otimes 5}, \ \ \  \cC=\{ \psi\in \cH^{\otimes 5}\, :\, S_j \psi =\psi \} $$
where $S_1 = X\otimes Z \otimes Z \otimes X \otimes I$, with 
$X,Y,Z$ the Pauli gates and $S_2,S_3,S_4,S_5=S_1S_2S_3S_4$ the cyclic permutations
of $S_1$, and with $\cH=\C^2$ the $1$-qbit Hilbert space. This is visualized as a code
over $\H^2$ that has one logical input at each tiles of the pentagon tessellation and
physical outputs across each edge of the tile, which are contracted with the legs
of the nearby tiles, so that the resulting holographic code has one logical input at
each tiles and outputs at the points at the boundary $\P^1(\R)$ that correspond to 
infinite sequences of tiles. 

\smallskip
\subsection{Triangle Fuchsian Groups and Holographic Codes}

In view of adapting this construction to the $p$-adic setting, it is better to
first consider a modification that will allow us to work directly with the
triangle Fuchsian group $\Gamma(2,4,5)$ rather than with its index $10$
subgroup $\Gamma$ of signature $(2,2,2,2,2)$ which is the symmetry
group of the regular right-angled pentagon tiling. 

\smallskip

This means replacing each pentagons in the tiling with its subdivision into
a triangulation of $10$ hyperbolic triangles with a vertex at the center of
the pentagon tile and the other vertices in the middle of the edges and
at the original vertices of the pentagon. 
\begin{center}
\includegraphics[scale=0.35]{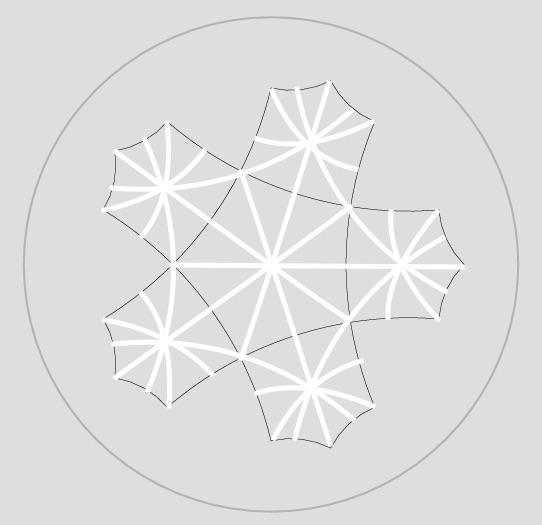} 
\end{center}
We then consider holographic codes
constructed by quantum codes associated to the triangle tiles. To this purpose,
we do not necessarily require the group to be $\Gamma(2,4,5)$. We can
work directly with the more general case of an arbitrary triangle Fuchsian group 
$\Gamma(a,b,c)\subset \PSL_2(\R)$ of hyperbolic type, $a^{-1}+b^{-1}+c^{-1}<1$. 

\smallskip

A simple way to construct a holographic code based on a tiling of the
hyperbolic plane realized by a hyperbolic triangle group is to use a 
quantum error correcting code described in \cite{HaPPY} that encodes
a single $3$-ary qubit (qutritt) into a space of three $3$-ary qubits by
$$ \begin{array}{rcl}
|0\rangle & \mapsto & |000\rangle + |111\rangle + |222\rangle \\
|1\rangle & \mapsto & |012\rangle + |120\rangle + |201\rangle \\
|2\rangle & \mapsto & |021\rangle + |102\rangle + |210\rangle\, .
\end{array} $$
This code can be represented as a perfect tensor $|a\rangle \mapsto T_{abcd} |bcd\rangle$
in the sense of \cite{HaPPY}. By placing a copy of this code (thought of as a copy of
the tensor $T_{abcd}$ at each triangle tile of the tiling specified by the Fuchsian triangle
group, one obtains a holographic code with a logical input qutritt at each tile
and physical output qutritts at points of the boundary $\P^1(\R)$ corresponding to
limit points of infinite sequences of tiles, from a specified base point in the bulk. 

\smallskip

A possible drawback of this simple construction is the fact that the
quantum code we are using does not contain any information about
the specific triangle group that determines the tessellation. This should
be corrected by taking into consideration the stabilizer subgroups of
edges and vertices, and incorporating them into the structure of the
quantum code. 

\smallskip

This can be done by considering quantum codes placed at the vertices,
rather than at the faces, of the tessellation of a hyperbolic triangle group 
$\Gamma(a,b,c)$. This requires using perfect tensors of different valences,
depending on the cardinality of the stabilizer group $G_v \subset \Gamma(a,b,c)$ 
of the vertex $v$. 

\smallskip

A triangle Fuchsian group $\Gamma(a,b,c)$ in $\PSL_2(\R)$ is generated
by elements $\gamma_1=\sigma_1 \sigma_2$, $\gamma_2=\sigma_2 \sigma_3$
and $\gamma_3=\sigma_3\sigma_1$, where the $\sigma_i$ with $\sigma_i^2=1$
are the reflections about the sides of the fundamental domain triangle in $\H^2$. 
The generators $\gamma_i$ satisfy the relations 
$\gamma_1^a=\gamma_2^b=\gamma_3^c=\gamma_1\gamma_2\gamma_3=1$,
that correspond to rotations by angles $2\pi/a$, $2\pi/b$ and $2\pi/c$, respectively,
with stabilizer groups $\Z/a\Z$, $\Z/b\Z$, $\Z/c\Z$ associated to the vertices of 
the tessellation. Let $\ell={\rm lcm}\{a,b,c\}$ and consider the embedding $\Z/a\Z\hookrightarrow \Z/\ell\Z$
by identifying $\Z/\ell\Z$ with $\ell$-th roots of unity and mapping the generator of $\Z/a\Z$ to
$\zeta^{\ell/a}$, where $\zeta$ is a primitive $\ell$-th root. Similarly, for the other two groups.
We can then consider a construction like the quantum codes described in \cite{HMSS}. At a 
vertex labelled by a stabilizer $\Z/a\Z$ we consider the polynomial code
$$ | \alpha \rangle \mapsto \sum_{\alpha_0,\ldots,\alpha_{a-1} \in \Z/\ell\Z} 
\otimes_{x\in \Z/a\Z} | f_{\underline{\alpha}}(x^{\ell/a}) \rangle $$
where $f_{\underline{\alpha}}(t)=\alpha_0 + \alpha_1 t + \cdots + \alpha_{a-1} t^{a-1}+\alpha t^a \in \Z/\ell\Z[t]$.
This encodes an input in $\ell^2(\Z/\ell\Z)$ into an output in $\ell^2(\Z/\ell\Z)\otimes \ell^2(\Z/a\Z)$, which
we think of as an $\ell$-ary qubit deposited at each side of the tessellation around the vertex. We can express
this as a tensor $T_{i_0\ldots i_a}$ with $a+1$ legs. By contracting legs along the matching edges of
the tessellations we obtain a holographic code that inputs an $\ell$-ary qubit at each vertex of the tessellation
and outputs at the points in the boundary $\P^1(\R)$ that are endpoints of geodesic lines consisting of edges
of the tessellation. 

\smallskip
\subsection{Surface Quantum Codes}

There is another interesting construction of quantum stabilizer codes associated to
tessellations of the hyperbolic planes, which was developed in \cite{Zemor}. These codes 
are constructed in general for a tiling defining a $2$-dimensional surface (possibly with
boundary). In particular, as shown in \cite{Zemor}, the construction applies to 
the case of hyperbolic triangle Fuchsian groups, through the associated Cayley graph
and the tessellation defined by it. In particular it applies to the triangle group 
$\Gamma(2,4,5)$ which we use here as a 
replacement for the right-angled pentagon tile of \cite{HaPPY}. The construction
of surface codes in \cite{Zemor} arises as a natural generalization of Kitaev's
toric code of \cite{Kitaev}. They have the advantage that they rely again on the CRSS
algorithm that coverts classical into quantum codes, hence they can be investigated
in terms of classical coding theory techniques. 

\smallskip

Consider a tessellation $\cR$ of a complex Riemann surface $\Sigma$ 
and its dual $\cR^*$ that has a vertex for each face of $\cR$
with two vertices being adjacent in $\cR^*$ if the corresponding faces in $\cR$ share a
common boundary edge. 
Let $\cE=(\epsilon_{v,e})$ be the vertex-edge incidence matrix
of $\cR$ and let $\cE^*$ be the vertex-edge incidence matrix of the dual graph $\cR^*$.
Let $V$ and $V^*$ be the $\F_q$-vector spaces spanned by the rows of $\cE$ and $\cE^*$, respectively.
The rows of $\cE$ are orthogonal to $V^*$ and the rows of $\cE^*$ are orthogonal to $V$, with
respect to the standard pairing $\langle v,v' \rangle =\sum_i v_i v'_i$.
The first homology groups of $\cR$ and $\cR^*$ can be identified with the quotients
$V^\perp/V^*$ and ${V^*}^\perp/V$. A quantum code can be associated to these data by
a version of the CRSS algorithm, using the pair of matrices $\cE$ and $\cE^*$. The
construction of the quantum code follows the same procedure illustrated above: to
pairs $(v,w)$ of vectors $v\in V$, $w\in V^*$, one associates an error operator $E_{(v,w)}$.
The condition that the spaces $V$ and $V^*$ are mutually orthogonal implies that
the bilinear pairing $\langle (v,w),(v',w')\rangle = \langle v,w' \rangle -\langle v',w\rangle$
vanishes, hence the group $\cS$ formed by these $E_{(v,w)}$ and the $\xi^j$, $0\leq j \leq p-1$
is abelian. Thus, one can associate to it a quantum stabilizer code by taking a common
eigenspace of the $E_{(v,w)}$. This imposes $\dim V + \dim V^*$ stabilizer conditions
on $n$ $q$-ary qubits, where $n$ is the number of columns of $\cE$ and $\cE^*$ (number
of edges of the graph $\cR$), hence the parameters of the resulting quantum code
are $[[n,k,d]]_q$, where $k=n -\dim V - \dim V^*$ and $d=\min\{ d_{V^\perp \smallsetminus V^*},
d_{{V^*}^\perp\smallsetminus V} \}$ with $d_{V^\perp \smallsetminus V^*}=\min\{ \omega(v)\,:\,
v\neq 0, \, v\in V^\perp \smallsetminus V^* \}$ with $\omega(v)=\#\{ i\,:\, v_i\neq 0 \}$ and
similarly for $d_{{V^*}^\perp\smallsetminus V}$. 

\smallskip

The Kitaev toric code consists of this construction applied to a graph $\cR$ obtained
by a tessellation of a torus into squares. Generalizations to other Riemann surfaces 
and other tessellations were described in \cite{Zemor}. The main idea is to associate
quantum surface codes to increasingly large portions of a given tessellation of the
hyperbolic plane or to suitable quotients of such regions.

\smallskip

In our case, we can start with the right-angled pentagon tessellation $\cR$
and its dual graph $\cR^*$. After choosing a root vertex $v_0$ of $\cR^*$ (the center
of a chosen face in the tiling) we denote by $\cR_N$ and $\cR_N^*$ the finite tessellations 
obtained by considering only the points that are up to $N$ steps away from $v_0$
(that is, such that the hyperbolic geodesic to $v_0$ passes through at most $N$ tiles. 
Let $V_N$ and $E_N$ be the number of vertices and edges in $\cR_N$ and let
$V_N^*$ be the number of vertices in $\cR_N^*$. 
The region $\cR_N$ has boundary, so in the construction of the dual
graph $\cR_N^*$ we assume that the dual graph has $E_N=E_N^*$ where the
edges of $\cR_N^*$ include an edge cutting through each boundary edge of $\cR_N$
and number of vertices $V_N^*$ given by the number of faces of $\cR_N$ plus
one additional vertex for each boundary edge of $\cR_N$. This will correctly produce,
in the limit when $N\to \infty$ boundary vertices on $\P^1(\R)$ at the endpoints of all 
geodesics of the dual graph $\cR^*$ of the tessellation $\cR$, which should be the 
physical outputs of a holographic quantum code. Note that, starting from the central
pentagon as zeroth step, at the first step one adds $10$ new pentagons, five
of which share an edge with the initial one and five that share a vertex. At the second
step, one adds $40$ new pentagons, where each of the $5$ pentagons of the first step
that shared an edge with the central pentagon (we call these tiles of the first kind)
will be adjacent to $2$ new tiles of the first kind (sharing an edge) and $1$ tile
of the second kind (sharing a vertex), while each of the $5$ tiles of the second kind will 
be adjacent to $3$ new tiles of the first kind and $2$ new tiles of the second kind.
Thus, if we let $F_N$ be the number of new tiles (faces) added to the tessellation at the $N$-th step,
with $F_N=m_N+n_N$, where $m_N$ and $n_N$ are, respectively, the number of tiles of the
first and second kind, namely those that share a full edge or just a vertex with a tile of the 
$(N-1)$-st step. We then have the recursion relation 
$$ m_{N+1} = 2m_N + 3 n_N , \ \ \  n_{N+1} = m_N + 2 n_N $$
with initial condition $m_1=n_1=5$. 
This gives
$$ (m_1,n_1)=(5,5), \ \  (m_2,n_2)=(25, 15), \ \ 
(m_3,n_3)=(95,55),  $$ 
$$  (m_4,n_4)=(355,205), \ \ 
(m_5,n_5)=(1325,765), \ \ 
(m_6,n_6)=(4945,2855), \ldots $$
which corresponds to $F_1=10$, $F_2=40$, $F_3=150$, $F_4=560$, $F_5=2090$, $F_6=7800\ldots$
Similarly, let $V_N$ denote the number of vertices added to the tessellation at the $N$-th step
in the construction. We count as before the numbers $m_N$ and $n_N$ of faces added at the $N$-th
step, and for each face we count new vertices counterclockwise, counting the leftmost vertex
(common to the next adjacent face) and not counting the rightmost vertex (which we include
in the counting for the next tile). This gives a number of new vertices equal to $W_N=2 m_N + 3 n_N =m_{N+1}$,
which is again computed in terms of the recursion above. We have $V_N = \sum_{k=0}^N W_k$
and $V_N^* = \sum_{k=0}^N F_k + E_{\partial,N}$, where $E_{\partial, N}$ is the number
of boundary edges at the $N$-th stage in the construction. This number is also equal to
$E_{\partial, N}=  2 m_N + 3 n_N=m_{N+1}$. 

\smallskip

One can also consider closed surfaces (without boundary) and associated quantum codes by passing to
Cayley graphs of quotient groups of the triangle Fuchsian group associated to
the tessellation.
In particular, the case that corresponds to the right-angled pentagon tile of the pentagon
code of \cite{HaPPY} is $m=4$ and $\ell=5$, for which we use the presentation
$$ \Gamma(2,4,5)=\langle a,b \,|\, a^2 =1, b^5=1, (ab)^4 =1 \rangle. $$
The $2$-complex used for the construction of the surface code in \cite{Zemor} 
is built by considering $2$-cycles of length 
$\ell=5$ and $2m=8$ of the form $\{ x,xb,xb^2,xb^3, xb^4,xb^5=x \}$ and $\{ x, xa, xab, xaba, x(ab)^2, x(ab)^2a, 
x(ab)^3, a(ab)^3a, x(ab)^4=x\}$, at every vertex $x$, where all vertices have valence $3$, with two 
edges $\{ x,xb \}$ and $\{x, xb^{-1} \}$ along an $\ell=5$-face and the remaining edge $\{x,xa\}$ along
a $2m=8$-face. By constructing an explicit matrix representation of $\Gamma(2,m,\ell)$ in the matrix
group $\SL_3(\Z[\xi])$, with $\xi=2\cos(\pi/m\ell)$,  and taking reduction of the matrix entries modulo a
prime $p$, one obtains a finite quotient group $G$, as the image of $\Gamma(2,m,\ell)$ (as a subgroup of
$\SL_3(\Z[\xi])$) in the quotient $\SL_3(\F_p[X]/(h(X)))$ where $h(X)$ is a function of the $2m\ell$-th
normalized Chebyshev polynomial. It is shown in \cite{Zemor} that this finite quotient group $G$ has the 
property that any word in the generators that is the identity in $G$ without being the identity in 
$\Gamma(2,m,\ell)$ must be of length at least $\log p$. This condition on the finite quotient group
ensures that the finite graph given by the Cayley graph of $G$ can be identified with a portion
of the infinite Cayley graph of $\Gamma(2,m,\ell)$, given by the neighborhood of
size $\log p$ of a vertex. Provided that $\log p$ is sufficiently large, the $2$-cycles will then correspond
to the $\ell$-cycles and $2m$-cycles in this region, as the only words within that length that are
equal to the identity in $G$ are those already equal to the identity in the triangle group.
The quantum code associated to the Cayley graph of $G$ and its dual graph then has code
parameters $[[n,k,d]]_q$ with $n=E$, dimension $k\geq \frac{E}{3}(1-2(\frac{1}{\ell}+\frac{1}{m})$, 
where $E$ and $V$ are the number of edges and vertices in the Cayley graph of $G$. Thus, the dimension 
grows linearly in the length of the code, while as shown in \S 3.3 of \cite{Zemor}, the 
minimum distance is proportional to $\log p$. 

\smallskip

The advantage of thinking in terms of triangle groups rather than pentagon codes is
that there is a parallel theory of $p$-adic hyperbolic triangle groups in $\PGL_2(\K)$, for 
$\K$ a (sufficiently large) finite extension of $\Q_p$, see \cite{Kato2}, \cite{Kato3}.
These are much more severely constrained than the Fuchsian triangle groups in $\PSL_2(\R)$
and only exist for small values of $p$.

\smallskip
\subsection{Triangle Groups on the Bruhat-Tits trees}

In order to consider analogous constructions in the Drinfeld $p$-adic upper half
plane $\Omega=\P^1(\C_p)\smallsetminus \P^1(\K)$, we first need to consider 
possible tilings of $\Omega$. As in the case of the real hyperbolic plane $\H^2$,
we can think of a tessellation of the Drinfeld plane $\Omega$ as a fundamental
domain $\cF$ for the action of a subgroup $\Gamma \subset \GL_2(\K)$ and
its translates $\gamma(\cF)$, $\gamma\in \Gamma$, with
the property that $\Omega/\Gamma$ is compact. Using the reduction map
$\Upsilon: \Omega \to \cT_\K$ from the Drinfeld plane to the Bruhat--Tits tree,
the property that $\Omega/\Gamma$ is compact translates into the property
that $\cT_\K/\Gamma$ is a finite graph. 

\smallskip

Unlike what happens in the case of Fuchsian groups acting on the real hyperbolic plane,
the existence of $p$-adic triangle graphs is much more severely constrained. One is
particularly interested in triangle groups of Mumford type. These are triangle groups
$\Gamma\subset \GL_2(\Q_p)$ such that 
$(\P^1(\C_p)\smallsetminus \Lambda_\Gamma)/\Gamma\simeq \P^1(\C_p)$ and
the uniformization map $\pi: \P^1(\C_p)\smallsetminus \Lambda_\Gamma\to \P^1(\C_p)$
is ramified at three points. 
In particular, by the classification result of \cite{Kato2}, \cite{Kato3} a $p$-adic triangle group 
of Mumford type, of signature $(2,4,5)$ exists only when $p=2$. No hyperbolic 
triangle groups of Mumford type exist for $p>5$. The complete list of hyperbolic $p$-adic
triangle groups $\Gamma(a,b,c)$ of Mumford type that can exist in the cases 
$p=2$, $p=3$, and $p=5$ is given in \cite{Kato3}.

\smallskip

Let $\cF$ be a fundamental domain for the action of the triangle group
$\Gamma(2,4,5)$ on the Drinfeld $p$-adic upper half plane $\Omega$
with $p=2$ and let $T$ be a fundamental domain for the action of the
same group $\Gamma(2,4,5)$ on the Bruhat--Tits tree $\cT_{\K}$ of a
(sufficiently large) finite extension $\K$ of $\Q_2$.
Since the reduction map $\Upsilon: \Omega \to \cT_{\K}$ is
equivariant with respect to the action of $\GL_2(\K)$, we can
assume that $T=\Upsilon(\cF)$. More generally, we can consider any
choice of one of the possible hyperbolic $p$-adic
triangle groups $\Gamma(a,b,c)$ of Mumford type, with $p\in \{2,3,5\}$,
acting on the Bruhat--Tits tree of a (sufficiently large) finite extension $\K$
of $\Q_p$, for one of these three possible values of $p$, 
and we proceed in the same way. 

\smallskip

A good way of describing the fundamental domain of the action of a finitely generated 
discrete subgroup $\Gamma \subset \PGL_2(\K)$ on the Bruhat--Tits
tree $\cT_\K$ and the resulting quotient graph is in terms of {\em graphs of
groups}, as shown in \cite{Kato2}, \cite{Kato3}. The theory of graphs of groups
was developed in \cite{Bass}, \cite{Serre}. A graph of groups consists of a finite 
directed graph with groups $G_v$ and $G_e$ associated to the vertices and edges 
of the graph, with $G_{\bar e}=G_e$, together with injective group homomorphisms 
$\varphi_s: G_e \to G_{s(e)}$ and $\varphi_t: G_e \to G_{t(e)}$
from the group associated to an edge to the groups associated to the source and target vertices. 
The fundamental group of a graph of groups is constructed choosing a spanning tree of the
graph: it is generated by the vertex groups $G_v$
together with an element $h_e$ for each edge $e$, with relations $h_{\bar e}=h_e^{-1}$ and
$$ h_e^{-1} \varphi_s(g) h_e = \varphi_t(g), \,\, \forall g\in G_e $$
and with $h_e=1$ for all $e$ in the chosen spanning tree. If one denotes by $\cG$ the graph
and by $G_\bullet$ the collection of groups associated to the vertices 
and edges, one writes $\pi_1(\cG,G_\bullet)=\varinjlim_{\varphi, \cG} G_\bullet$ for the resulting
amalgam given by the fundamental group of the graph of groups. In the case where the graph
consists of one edge and two vertices, this fundamental group is just the pushfoward in
the category of groups, namely the amalgamated free product $G_{s(e)}\star_{G_e} G_{t(e)}$.
The main idea (see \cite{Bass}, \cite{Serre}) is to associate to the action of a discrete
group on a tree a quotient given not just by a graph but by the richer structure
of a graph of groups, which keeps track of the information about the stabilizers of vertices
and edges. In the case of a discrete subgroup $\Gamma \subset \PGL_2(\K)$, we consider
the tree of groups given by the subtree $\cT_\Gamma$ of the Bruhat--Tits tree $\cT_\K$
together with the stabilizers $G_v$ and $G_e$ of vertices and edges, and we obtains a
graph of groups as the quotient graph $\cT_\Gamma/\Gamma$. 
It is shown in \cite{Kato3} that $p$-adic triangle groups of Mumford
type are characterized by the property that the quotient graph $T=\cT_\Gamma/\Gamma$ is
a tree consisting of three lines meeting at a single root vertex $v_0$. Such trees are called {\em tripods}. 
This tree, decorated
with the stabilizer groups of vertices and edges is a tree of groups. The ends of this tree
are the three branch points, at $0$, $1$ and $\infty$, 
of the genus zero curve $\Omega_\Gamma/\Gamma$. The group $\Gamma$
can be reconstructed from the tree of groups $(T, G_\bullet)$ as the associated fundamental 
group, \cite{Kato2}. Indeed the possible
$p$-adic triangle groups of Mumford type are explicitly constructed using this method. For
example, the tripod associated to the $p$-adic triangle group $\Gamma(2,4,5)$ with $p=2$,
seen as a tree of groups, is the case $\ell=m=1$ of the following family (from \cite{Kato3}):
\begin{center}
\includegraphics[scale=0.35]{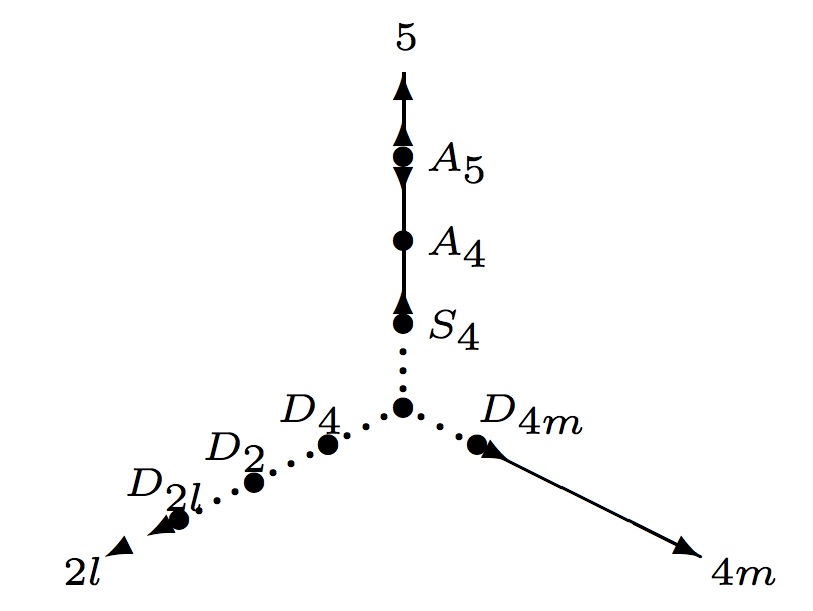} 
\end{center}
with subgroups $D_2 \subset D_4 \subset S_4$ and $D_2$ intersecting $A_4\subset S_4$ trivially.
In the case $\ell=m=1$ the resulting amalgam agrees with the pushout $S_4\star_{A_4} A_5$.

\smallskip
\subsection{Tessellations of the Drinfeld Plane}

A general algorithm exists for computing fundamental domains in Bruhat--Tits trees
for the action of certain quaternion groups, see \cite{FraMas}. In these cases the algorithm
produces 
\begin{enumerate}
\item a connected subtree $\cD_\Gamma$ of the Bruhat--Tits tree which is a fundamental domain
for the group action, in the sense that the edges of $\cD_\Gamma$ form a complete set of
coset representatives for $E(\cT)/\Gamma$;
\item the edge and vertex stabilizer groups $G_e$, $G_v$ for $e\in E(\cD_\Gamma)$ and
$v\in V(\cD_\Gamma)$;
\item an explicit form for the quotient map by identifications $(v,v',\gamma)$ between
pairs of boundary vertices $v,v'$ of the fundamental domain $\cD_\Gamma$, with 
$\gamma\in \Gamma$ such that $v'=\gamma v$.
\end{enumerate}

\smallskip

This algorithm can be used to produce corresponding tessellations of the Drinfeld
$p$-adic upper half plane. Let $\Gamma$, $\cD_\Gamma$, $G_e$, $G_v$, and $\{ (v,v',\gamma) \}$
be given as above, through the algorithm of \cite{FraMas}.
Using the projection map $\Upsilon: \Omega \to \cT$ from the Drinfeld plane to the Bruhat--Tits tree, 
we can construct an associated tessellation of the Drinfeld plane, where the tiles are given by $\gamma T$,
with $\gamma \in \Gamma$ and 
$$ T = \bigcup_{v\in V(\cD_\Gamma)} \Upsilon^{-1}(v) \cup \bigcup_{e\in E(\cD_\Gamma)} \Upsilon^{-1}(e). $$
The gluing rules for the tiles are prescribed by the data (2) and (3) associated to the fundamental
domain on the Bruhat--Tits tree.

\smallskip
\subsection{Lifting Holographic Codes from the Bruhat--Tits Tree}\label{LiftSec}

Another way to obtain holographic codes on the Drinfeld plane is to
lift the construction of the classical and quantum codes on Bruhat-Tits trees
described in \S \ref{BTcodes} via the surjection $\Upsilon: \Omega \to \cT_\K$. 
This means that the ``tiles" to which we associate classical and quantum codes
in the Drinfeld plane are given, in this case, by the regions $\Upsilon^{-1}(v)$,
the preimages in $\Omega$ of vertices of the Bruhat--Tits tree, and the outputs
of each (classical or quantum) Reed--Solomon code is stored in the connecting
regions $\Upsilon^{-1}(e)$. This can be done by choosing a lift of the
projection $\Upsilon$, which realizes the Bruhat--Tits tree as a skeleton of $\Omega$
and constructing the holographic code over that skeleton. The choice of a lift of the
projection is non-canonical, hence this type of construction has the same kind of drawback
of the construction used in \cite{HMSS} to simulate the pentagon code via a
choice of a planar embedding of a tree along edges of the pentagon tiling of the
real hyperbolic plane. An advantage in this case, however, is that the projection
$\Upsilon$ is equivariant with respect to the $\GL_2(\Q_p)$ symmetries so one
maintains the symmetries of the tree intact, unlike the case of the
planar embedding used in \cite{HMSS}.

\medskip
\section{Holographic Codes on Higher Rank Bruhat--Tits Buildings}\label{BuildCodes}

As above, we denote by $\cT_{n,\K}$ the Bruhat--Tits building of $\GL_{n+1}(\K)$
and by $\Omega_n$ the Drinfeld symmetric space.

\smallskip

Consider first the case of the Bruhat--Tits building of $\GL_3(\K)$, with $\K$ a finite
extension of $\Q_p$ with residue field $\F_q$, $q=p^r$. The set of vertices
adjacent to a given vertex $v\in V(\cT_{2,\K})$ is a bipartite set, consisting of
the set of $q^2+q+1$ $\F_q$-rational points of the projective plane $\P^2$ over $\F_q$ 
together with the set of $q^2+q+1$ $\F_q$-rational lines of the projective plane
$\P^2$ over $\F_q$. The surface $X$ over $\F_q$ obtained by blowing up all the 
$\F_q$-rational points of $\P^2$ contains an exceptional divisor (a line) for
each $\F_q$-rational points of $\P^2$ and a proper transform (also a line) for
each $\F_q$-rational line in $\P^2$. Thus, to each vertex $w$ adjacent to the
given vertex $v$ we associate a line $\ell_w$ in the blowup surface $X$. 
Let $u,w$ be vertices adjacent to $v$: the set $\{ u,v,w \}$ corresponds to a 
$2$-simplex in the $2$-dimensional simplicial complex $\cT_{2,\K}$ if and 
only if the lines $\ell_u$ and $\ell_w$ intersect nontrivially in $X$. 

\smallskip

In the case of $\Q_2$ one obtains the well known picture below, with the $7$ points and $7$ lines
of $\P^2(\F_2)$ as vertices and with $21$ edges, \cite{Cox}.
\begin{center}
\includegraphics[scale=0.35]{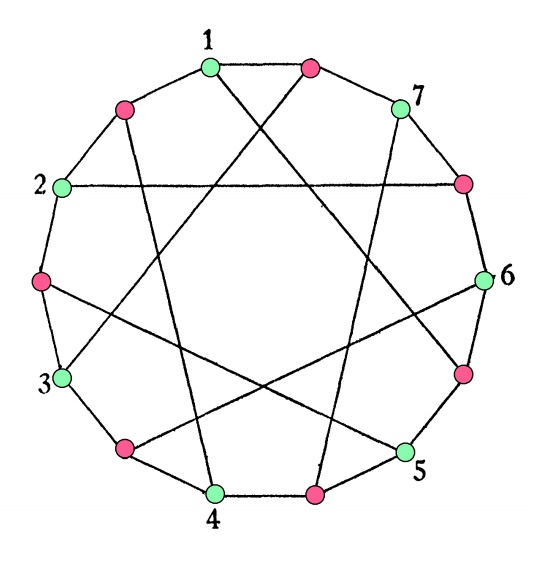} 
\end{center}

In order to extend the construction of holographic codes to higher rank Bruhat--Tits
buildings, in a way that reflects the associated geometries over finite fields that
determine the local structure of the building, we need to replace the classical Reed--Solomon
codes with algebro-geometric codes associated to higher-dimensional algebraic varieties.

\smallskip
\subsection{Codes on the Bruhat--Tits buildings of $\GL_3$ from algebro-geometric codes on
surfaces}

A general procedure for constructing algebro-geometric codes over higher-dimensional
algebraic varieties generalizing the Reed--Solomon codes is described in \cite{TsfaVla},
see also \cite{Hansen}. Given a smooth projective variety $X$ over $\F_q$ with an ample
line bundle $\cL$, one obtains a linear code $C(X,\cL,\cP)$, where $\cP$ is a set of
$\F_q$-algebraic points of $X$, as the image of the germ map
$$ \alpha: \Gamma(X,\cL) \to \oplus_{x\in \cP} \cL_x \simeq \F_q^n, $$
which evaluates sections $s\in \Gamma(X,\cL)$ at points $x\in \cP$, with the last
identification given by a choice of an isomorphism $\cL_x \simeq \F_q$ of the fibers
at $x\in \cP$, with $n=\# \cP$. 

\smallskip

For example, for $X=\P^2$, with $\cL=\cO(m)$, with $0<m\leq q$, and $\cP$ the set of all 
$\F_q$-rational points of $\P^2$, one obtains a code $C(\P^2, \cO(m), \P^2(\F_q))$
with length $n=q^2+q+1$, dimension $k=\frac{1}{2}(m+1)(m+2)$, and minimum
distance bounded by $d\geq q^2+q+1 - m(q+1)$, see \cite{Hansen}.

\smallskip

We focus here on the case of the Bruhat--Tits building of $\GL_3(\K)$, with $\K$ a finite
extension of $\Q_p$ with residue field $\F_q$, $q=p^r$. As we mentioned above, the link
of a vertex in the Bruhat--Tits building is described in terms of the geometry of an
algebraic surface $X$ obtained by blowing up all the $\F_q$-algebraic points of $\P^2$.

\smallskip

We use the example above of algebro-geometric codes $C(\P^2, \cL, \P^2(\F_q))$
associated to line bundles $\cL$ over $\P^2$ to construct a classical holographic code on
the Bruhat--Tits building of $\GL_3(\K)$. We fix a base vertex in the building and assign 
as logical input the datum of a divisor $D$ on $\P^2$ so that $\cL=\cL(D)$.
Consider then the surface $X$ over $\F_q$ obtained by blowing up all the $\F_q$-rational
points of $\P^2$, and the pullback $\pi^*\cL$ under the projection map, and line bundles
of the form $\hat\cL=\pi^*\cL \otimes \cO(-\sum_i k_i E_i)$ where the $E_i$ are the exceptional
divisors of the blowup. Assume that $D$ and the $k_i$ are chosen so that $\hat\cL$
is represented by an effective divisor on $X$.
We now consider the $q^2+q+1$ lines in $X$ determined by the $\F_q$-lines of $\P^2$
and the $q^2+q+1$ lines that correspond to the $\F_q$-points of $\P^2$ and the
set $\cP$ consisting of the $q+1$ $\F_q$-rational points of each of these lines, with
$\# \cP = 2 (q+1) (q^2+q+1)$. The code $C(X,\hat\cL, \cP)$ can be viewed as a code
that, given the logical input $D$ at the base vertex $v$, deposits an output given by a 
vector in $\F_q^{q+1}$ at each adjacent vertex $w$ in the Bruhat--Tits building. These
outputs are related by a consistency condition, which is determined by the edges and
$2$-cells of the building. Namely, whenever $w$ and $u$ are vertices adjacent to $v$,
such that $\{v,w,u \}$ is a $2$-cell in the building, we know the corresponding condition
on $X$ is that the two lines $\ell_w$ and $\ell_u$ intersect. The presence of a point
of intersection means that the corresponding vectors in $\F_q^{q+1}$ must agree in one of 
the $q+1$ coordinates. 

\smallskip

When one propagates the construction to nearby vertices in the Bruhat--Tits building,
part of the logical input is reserved for the output $\F_q^{q+1}$-vector of the nearby
vertices already reached by the previous steps from the chosen root vertex. As in the
case of the Bruhat--Tits tree, we identify the given $\F_q^{q+1}$-vector (computed 
as output by the previous code) with assigned values at one of the lines in $X$ that 
corresponds to one of the lines in $\P^2$ (which we can think of as the $\P^1$ 
at infinity in $\P^2$). There is a consistency condition for the output at a new vertex $w$
that is adjacent to a $2$-cell where the remaining two vertices $v$ and $v'$ already have 
outputs $\underline{x}(v), \underline{x}(v')\in \F_q^{q+1}$ assigned by the previous codes: 
the outputs $\underline{x}(v), \underline{x}(v')$ at the two previous vertices $v,v'$
are two vectors in $\F_q^{q+1}$ that agree in one coordinate, hence they
fix the values of the sections at two intersecting lines in $X$. The resulting output $\underline{x}(w)$  
at the new vertex $w$ is then computed by the values at the $q+1$ points of the line $\ell_w$
of all sections $s$ that satisfy the constraints given by the assigned values at the points
of $\ell_v$ and $\ell_{v'}$. The construction can in this way be propagated to the rest of
the Bruhat--Tits building of $\GL_3(\K)$. This illustrates the general approach to
constructing classical holographic codes on higher rank Bruhat--Tits buildings.

\smallskip

A construction of quantum holographic codes can be obtained from these classical  
codes using a version of the CRSS algorithm (possibly by allowing more general
types of weighted versions of the classical codes, as we discussed in the case
of the Reed--Solomon codes). The details of the corresponding quantum codes
for higher rank buildings will be discussed in forthcoming work. 

\smallskip
\subsection{Codes on Drinfeld symmetric spaces}

Another possible approach to the construction of holographic codes 
for higher-rank $p$-adic symmetric spaces consists of working with
Dirnfeld symmetric spaces instead of Bruhat--Tits buildings. This 
extends the approach discussed in \S \ref{DplaneSec} on codes
associated to actions of discrete groups on the Drinfeld plane. 

\smallskip

In the higher rank setting, we consider two possible viewpoints.
The first is based on the projection map from the Drinfeld
symmetric space $\Upsilon: \Omega_n \to \cT_{n,\K}$, from the
Drinfeld space $\Omega_n = \P^n(\C_p)\smallsetminus \cup_{H\in \cH_\K} H$
(the complement of the $\K$-rational hyperplanes in $\P^n$) to the
Bruhat--Tits building of $\GL_n(\K)$. The idea here, as in 
\S \ref{LiftSec} above, is to lift via the projection map a construction
of holographic classical and quantum codes from the Bruhat--Tits
building to the space $\Omega_n$, with logical inputs associated
to the regions $\Upsilon^{-1}(v)$, with $v$ the vertices of $\cT_{n,\K}$
and outputs and compatibility conditions along the edges, faces, and 
higher-dimensional cells. Since the projection map $\Upsilon$ is
equivariant with respect to the $\GL_n(\K)$ action, whatever symmetry
the codes constructed on $\cT_{n,\K}$ exhibit will be inherited
by the resulting codes on $\Omega_n$. 

\smallskip

The other possible approach consists of constructing a tensor network 
directly associated to a given action of a discrete subgroup $\Gamma$ of
$\GL_n(\K)$ on the symmetric space $\Omega_n$. Roughly, the main idea
in this case is to assign logical inputs to the fundamental domains of the
action, while outputs should be associated to the generators of the discrete group with
compatibility conditions resulting from the relations. In this way, the codes
assigned to each copy of the fundamental domain can be compatibly
assembled into a global holographic code on $\Omega_n$, with logical
inputs in the bulk and outputs at the boundary.  The outputs should live on the
points in the limit set of the group action on the rational hyperplanes
$H\in \cH_\K$.  We will discuss these constructions of holographic codes
on higher rank $p$-adic symmetric spaces in forthcoming work.

\bigskip
\bigskip

\subsection*{Acknowledgment} The author thanks Matthew Heydeman, Sarthak Parikh, and
Ingmar Saberi for many very useful discussions and an ongoing collaboration on
several topics discussed in this paper, and especially Sarthak Parikh for suggesting
several improvements to the paper. The author is partially supported by NSF grant 
DMS-1707882 and by the Perimeter Institute for Theoretical Physics.

\end{document}